\documentstyle[aps,twocolumn,prl]{revtex}

\begin{document}
 
\newcommand{\lsim}{\raisebox{1.5pt}{\small $<$}\hspace*{-6.7pt}\raisebox{-
3pt}{\small $\sim$} }
\newcommand{\gsim}{\raisebox{1.5pt}{\small $>$}\hspace*{-6.7pt}\raisebox{-
3pt}{\small $\sim$} }

\draft

\title{Thermal relaxation in charge ordered 
\mbox{Pr$_{0.63}$ Ca$_{0.37}$MnO$_{3}$}  in presence of 
a \mbox{magnetic field}} 
\author{A.K.Raychaudhuri and Ayan Guha}
\address{Department of Physics, Indian Institute of Science, 
Bangalore 560 012, India}
\author{I.Das and R. Rawat}
\address{Inter University Consortium for DAE facilities, University Campus,
Khandwa Road, \mbox{Indore- 452017, India}}
\author{C.N.R. Rao}
\address{CSIR Center of Excellence in Chemistry, Jawaharlal Nehru Center for
Advanced Scientific Research, Jakkur P.O., \mbox{Bangalore 560 064,} India}
\date{\today}

\twocolumn[\hsize\textwidth\columnwidth\hsize\csname 
@twocolumnfalse\endcsname

\maketitle
\begin{abstract}
We report observation of substantial thermal relaxation in 
single crystal of charge ordered system Pr$_{0.63}$Ca$_{0.37}$MnO$_3$
in an applied magnetic field of H = 8T. The relaxation is observed when 
the temperature is scanned in presence of a magnetic field in the temperature 
interval $T_{MH}<T<T_{CO}$ where $T_{CO}$ is the charge ordering temperature 
and $T_{MH}$ is charge melting temperature in a field. In this temperature 
range the system has coexisting charged ordered insulator (COI) and 
ferromagnetic metallic (FMM) phases. No such relaxation is observed in the 
COI state in H = 0T or in the FMM phase at $T < T_{MH}$ in presence of 
a magnetic field. We conclude that the thermal relaxation is due to 
two coexisting phases with nearly same free energies but separated by a 
potential barrier. This barrier makes the transformation from one phase to 
the other time-dependent in the scale of the specific heat experiment and 
gives rise to the thermal relaxation. 

\end{abstract}

]

Rare earth manganites with general chemical formula Re$_{1-x}$Ae$_{x}$MnO$_{3}$, 
has attracted widespread interest because of a variety of fascinating properties like 
colossal magnetoresistance (CMR) and charge ordering (CO) ~\cite{Kuwahara,Raorev}. 
For certain values of x close to 0.5, these compounds undergo a first order
transition at a temperature $T_{CO}$, where the Mn$^{3+}$ and Mn$^{4+}$ species 
arrange themselves in a commensurate order in the lattice. The charge-ordered
state is an insulating state.  An interesting aspect of this charge ordered 
insulating (COI) state (which is also accompanied by orbital ordering) is that it 
is unstable under a number of external perturbations like electric and magnetic
fields, x-rays, optical radiation etc. In presence of an applied magnetic field, 
the sample undergoes an insulator- metal transition (generally termed as``melting") 
from a COI state to a ferromagnetic metal (FMM) state. In presence of magnetic
field, the temperature where this melting occurs is termed as $T_{MH}$. 

In this paper our particular interest is the region between $T_{MH} < T < T_{CO}$ 
in presence of an applied magnetic field. Evidence is now gathering that in 
this region (which we call mixed charge ordered region, (MCO)) the two phases, 
namely the charge ordered insulating (COI)phase and the spin ordered ferromagnetic 
metallic (FMM) phase coexist. This coexistence is due to electronic phase 
separation between two phases of nearly the same free energy. Recent transport 
(including non-linear transport), noise, magnetization experiments etc. have shown 
that in this class of manganites electronic phase separation is a very commonly 
occuring phenomenon ~\cite{Podz,Ayan1,Ayan2,Dagotto,Fath}. In this communication 
we report an interesting observation that in the region 
of coexisting phases (which has been created by applying a magnetic field below 
$T_{CO}$)  large thermal relaxation effects arise during the measurement of 
specific heat. From our experiments we concluded that though the two coexisting 
phases have nearly the same free energy (and nearly the same specific heat) they 
are separated by an energy barrier which makes the 
time of conversion of one phase to the other measurable in the time scale of 
specific heat experiments. This leads to the observation of thermal relaxation. 
We also correlate these measurements to observed hysteresis in the I-V curve 
and low energy conductance noise.

Present investigation was carried out on a single crystal of 
Pr$_{0.63}$Ca$_{0.37}$MnO$_3$ which has been grown by float zone technique and 
has been used in a number of previous experiments by our group particularly in 
investigation of non-linear transport studies in the MCO region and electric 
field induced instabilities.  Pr$_{1-x}$ Ca$_{x}$MnO$_{3}$ is a prototypical 
charge-ordered system, which because of its small average cation radius 
$< r_A >$ remains insulating for all values of x. This system is unique and 
perhaps the most studied among all the charge-ordered systems. The resistivity 
($\rho$) vs. T curve in zero field and in a field of H = 8T are shown in the
inset (b) of figure~1. Inset (a) of the same figure shows the H-T phase diagram
 of our sample. We have performed our experiments along the constant H = 8T line
shown by an arrow in the inset. The COI state can be melted to a FMM state by 
application of a magnetic field of 8T at $T_{MH} \approx$ 90K. 
This particular composition  shows both charge and orbital ordering below 
$T_{CO}$ = 235K and the antiferromagnetic (AFM) ordering occurs at $T_N$ = 170K. 

The thermal relaxation of the sample was measured using an semi-adiabatic 
calorimeter ~\cite{Idas} operating in the 20K $< T <$ 300K range in 0T and 
after field cooling in 8T. The temperature of the sample was monitored as a 
function of time after applying a known heat pulse. If the specific data are 
calculated after the sample has relaxed thermally (see discussion below) we 
can measure the equilibrium $C_p$. The heat capacity shown in figure 1 both 
for H = 0 and 8T are equilibrium specific heats. In the same graph the dotted 
line shows the estimated vibrational back ground. It can be seen that the two 
specific heats are nearly the same except near $T_{CO}$. The large peak near 
$T_{CO}$ is a consequence of latent heat in the first order phase transition at 
$T_{CO}$. A complete description of the equilibrium specific heats and the 
estimation of the vibrational background is given elsewhere ~\cite{Raych}. 
It is important to note that in the temperature range of interest (except the region 
$T \approx T_{CO}$), the vibrational background makes up for the bulk of the 
observed specific heat. The closeness of the $C_{p}$ for H = 0 and 8T implies 
that the coexisting phases in the mixed charge ordered region at H = 8T have 
energy similar to the COI phase existing at H = 0T.(Note: There is a small latent 
heat release ($\approx$ 10J/mole) when the FMM phase is heated to the COI phase 
at $T_{MH}$). The focus of this paper is the thermal relaxation experiment 
as described below.

In a typical measurement of heat capacity using adiabatic or semi adiabatic 
heat-pulse technique, a measured amount of thermal energy {\it Q} is 
applied to the sample producing a change in temperature $\Delta T$. The sample 
temperature ($T_s$) raised above the bath temperature ($T_o$) relaxes to its 
original value as described by the following equation :
\begin{equation}
\Delta~T(t) = (T_s - T_o)~exp(-t/\tau_B)
\end{equation}
where the time constant $\tau_B = C_p~/~K_w$, $K_w$ being the thermal 
conductance of the weak link that connects the sample to the bath. By isolating 
the sample from the bath (to satisfy the adiabatic condition), the relaxation 
time {\mathversion{bold}$\tau_B$} is made as long as experimentally convenient. 
Equation~1 assumes that $\tau_B >>\tau_I$, where $\tau_I$ is the internal thermal 
equilibrium time of the sample defined as $\tau_I =L^{2}/D_{th}$, $D_{th}$ is 
the thermal diffusivity of the sample of dimension L. As long as $\tau_B >>\tau_I$, 
the thermal relaxation of the sample after the application of the heat pulse can 
be described by a single relaxation time as in eqn. 1. If the sample has very 
low $D_{th}$ or has internal equilibrium taking place over a long time then 
the ratio $\tau_I/\tau_B$ becomes finite and the relaxation curve 
($\Delta T~- ~t$) cannot be described by a single exponential decay. 

In absence of magnetic field , we observe (both for $T>T_{CO}$ and $T<T_{CO}$)
that the sample thermally relaxes to the bath temperature with a long time 
constant, $\tau_B$, which is typically $\gg 500 secs$. In this case the internal 
relaxation time, $\tau_I~<~1 sec$ and a single relaxation time $\tau_B$ describes 
the thermal relaxation of the sample for all T .  

In an applied field of 8T, for $T > T_{CO}$ and $T < T_{MH}$ we observe that 
the thermal relaxation is still governed by a single relaxation time as in 
the H = 0T case (see fig. 2(a) and 2(c)). However, this breaks down for the 
MCO state ($T_{CO} > T > T_{MH}$) as shown in the ($\Delta T$ - t) curves in 
figure 2(b). The curve shows two relaxation time behavior, which occurs when 
$\tau_I$ is finite compared to  $\tau_B$. We have analyzed the observed  
thermal relaxation curve at a given T using the relation :
\begin{equation}
\Delta T = A~exp~{(-t/\tau_I)}~ + ~B~exp~{(-t/\tau_B)}
\end{equation}

where A and B are constants at a given T. From the analysis of the observed 
relaxation curves of the MCO state at different T we evaluated $\tau_I$ as 
a function of temperature and this is shown in figure~3(a). The value of 
$\tau_B$ is $\approx$ 500 secs to 1500 secs in the whole temperature range. 
In the MCO state, $\tau_I$ (H = 8T)is much larger than $\tau_I$ (H=0T) which 
is typically $\leq 1sec$. Interestingly, $\tau_I$ is finite and measurable 
only in the temperature interval $T_{MH} < T < T_{CO}$ in a magnetic field. 

Large value of $\tau_I$ is a manifestation that the sample reaches internal 
thermal equilibrium over an extended period and this makes a part of the 
specific heat relaxing or time dependent over the experimental time scale. 
We can obtain a quantity C$_{inst}$ immediately after the heat pulse is 
applied without waiting for complete thermal relaxation. This is the specific 
heat capacity in the scale $t ~\rightarrow$ 0 which contain contributions 
of only those constituents which relax faster than the experimental time 
scale $\tau_I$. This is less than the equilibrium specific heat $C_p$, which 
we measure after the sample has internally equilibrated ($t \gg \tau_I$). 
The difference of these two $\Delta C$ = C$_p$ - C$_{inst}$ gives a measure 
of the time dependent or relaxing  heat capacity.
We plot $\Delta C(8T)/C_p$ as a function of T in figure~3(b). For H = 0T,
$\Delta C$ = 0 whereas $\Delta C(8T)/C_p ~\sim~$ 25 \% in the range 
$T_{CO} > T > T_{MH}$. (There is a small contribution of $\Delta C$ 
persisting down to 50K, a temperature close to the spin canting temperature 
in zero magnetic field.)
 
Previous experiments of non-liner transport as well as noise in this system 
(and related systems) have led to the conclusion that the MCO comprises of 
coexisting CO insulating and charge-molten FM metallic phases, as has been 
stated earlier. The fact that we donot see the thermal relaxation when the 
sample is completely in COI phase ($T>T_{CO}$)or completely in FMM phase 
($T<T_{MH}$) clearly establish that the thermal relaxation arises from 
coexisting phases. We have analyzed the resistivity data (shown in fig.~1(b)) 
in the MCO region with the help of such a scenario. Using an effective medium 
theory ~\cite{Kirk} we find that a fraction ($f_m$) of the FMM phase grows 
steadily below $T_{CO}$, it reaches $f_m\approx 0.2$ at $T/T_{CO} \approx 0.8$ 
and the percolation volume $f_{m}=1/3$ at $T = T_{MH} \approx 90K$. We propose 
that the observed relaxation is arising because of the co-existing phases, 
whose relative volume fractions change as the temperature change on application 
of the heat pulse, have a finite time of conversion from one phase to the other. 
This has been elaborated below. In addition to this effect, close to  $T_{MH}$ 
and T$_{CO}$ additional relaxation arises due to the release of the latent heat 
(see figure 3(b)). 

It has been pointed out earlier that the observed equilibrium $C_{p}$ is 
nearly the same both for H = 0 and 8T and they are close to the vibrational 
background specific heat except the region close to $T_{CO}$. This would 
clearly imply that energetically the H = 0T CO phase and the H = 8T MCO 
phase (which a has a fair fraction of FM phase) are quite similar. If the 
coexisting phases in the sample would have been able to transform 
from one phase or to the other within a time scale  $<<$ 1sec then our 
calorimetry experiment would not have distinguished between these two phases. 
There would not have been any thermal relaxation of the type we have observed. 
The situation, however, changes when there is a finite time associated with 
the transformation from one phase to the other coexisting phase. At any given 
temperature, there is an average equilibrium relative fraction ($f_m$) of the 
two coexisting phases. The noise experiments on these materials show that at 
a given temperature the equilibrium volume fraction is result of  dynamic 
equilibrium and in the time scale of the noise experiments (typically $>$1 sec) 
the two phases can transform one from the other giving rise to the 
conductivity fluctuation. In a heat pulse experiment when we apply the pulse 
and the temperature changes, the equilibrium volume fraction also changes so 
that the volume fraction corresponds to the new temperature. The two phases 
though energetically similar are separated by an energy barrier which makes 
this change take place over a finite time. Thus the sample thermally relaxes 
over a longer period and the relaxation time $\tau_I$ becomes finite comparable 
to $\tau_B$.

Large resistive relaxation and delayed release of heat has been observed in 
similar Pr$_{1-x}$ Ca$_{x}$MnO$_{3}$ samples at lower temperature (typically for 
$T < 50K$) when the samples were cooled  to low temperature and the applied 
field was cycled at a fixed temperature ~\cite{Anane,Mroy}. The time scales 
observed at those temperature were large (in the scale of $10^3$ secs or more). 
In our experiment the temperatures are much higher ($T_{MH}<T<T_{CO}$) and they 
were carried out by keeping the field constant and varying the temperature. The 
thermal relaxation observed by us occurs in the scale of few tens of secs 
presumably because we are doing the experiments at higher T($>100K$). However, 
the underlying physics seems to be similar in these experiments.
 
To conclude, we have observed that when a magnetic field is applied to a CO 
solid in the range $T_{MH}<T<T_{CO}$ a large thermal relaxation sets in when 
a heat pulse is applied.The relaxation is not seen in the pure phases at 
$T<T_{MH}$ or $T>T_{CO}$ or at all T in H=0T. We explain this observation as 
arising from two coexisting phases (COI and FMM) of similar energy which are 
separated by a potential well, which imposes a finite time(measurable in the 
experimental time scale) of conversion of one phase to the other.

A.G. thanks the CSIR Center of Excellence in Chemistry, JNCASR, for financial
support. AKR thanks  DST for partial support through a sponsored scheme.

\newpage
{\bf Figure captions :}
\vspace{0.5cm}

{\bf figure 1:} Equilibrium specific heats $C_p$ of Pr$_{0.63}$
Ca$_{0.37}$MnO$_3$ single crystal in presence of 0 and 8T magnetic fields.
(data from ref. ) Inset (a) shows the H-T phase diagram and the arrow shows 
the path taken for the experiment. Inset (b) shows $\rho$ vs. T for H = 0 and 8T.

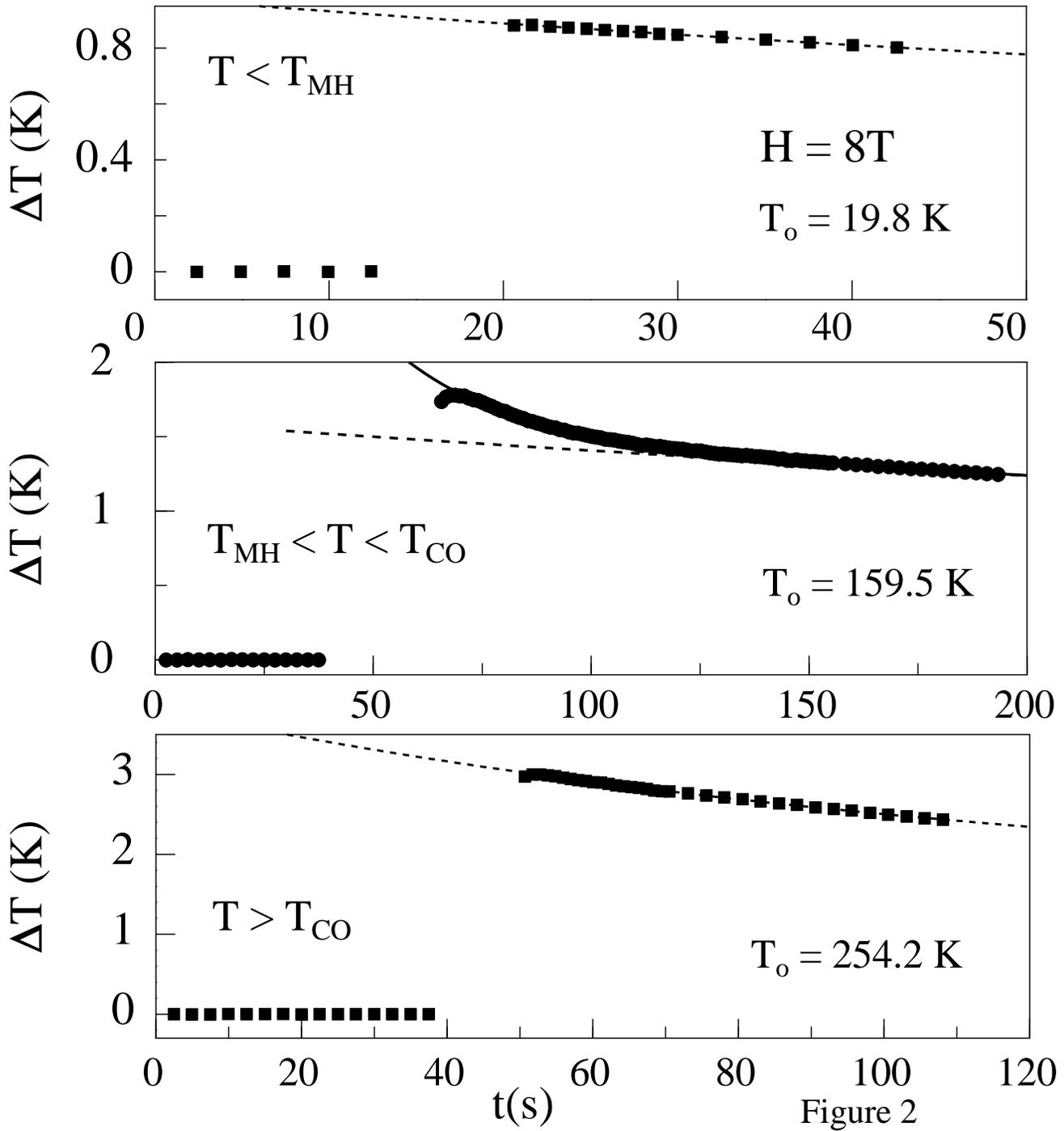
{\bf figure 2:} Thermal relaxation curves ($\Delta$T - t) of the sample in 
presence of H = 8T for three temperatures $T < T_{MH}$, $T_{MH} < T < T_{CO}$
and $T > T_{CO}$. Note the two relaxation time behaviour for $T_{MH} < T < 
T_{CO}$.

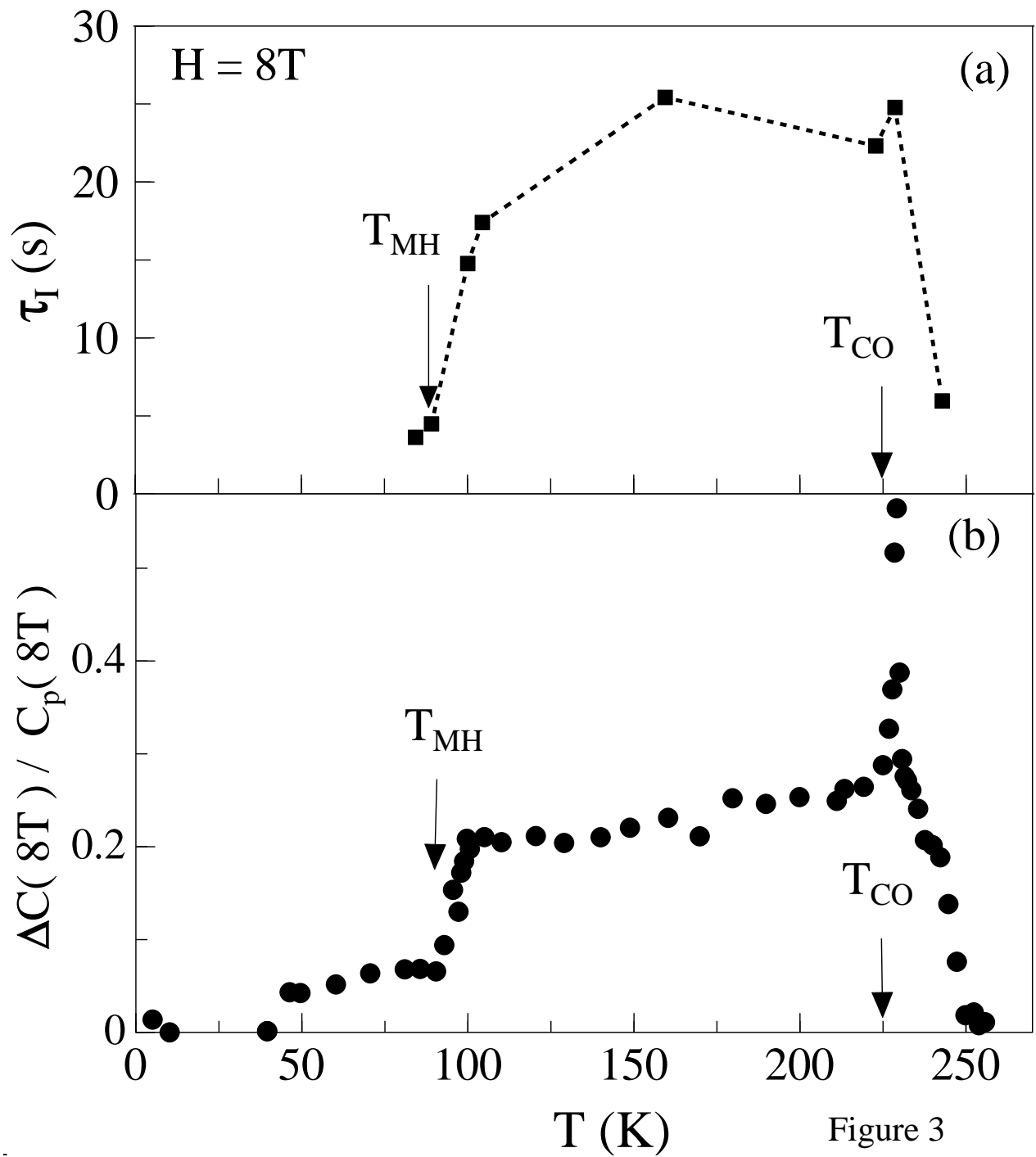
{\bf figure 3:} (a) Variation of sample thermal relaxation time $\tau_I$
with $T$. (b) Variation of relaxing $C_p$ with $T$ (see text). The peak
near $T_{CO}$ is due to latent heat.

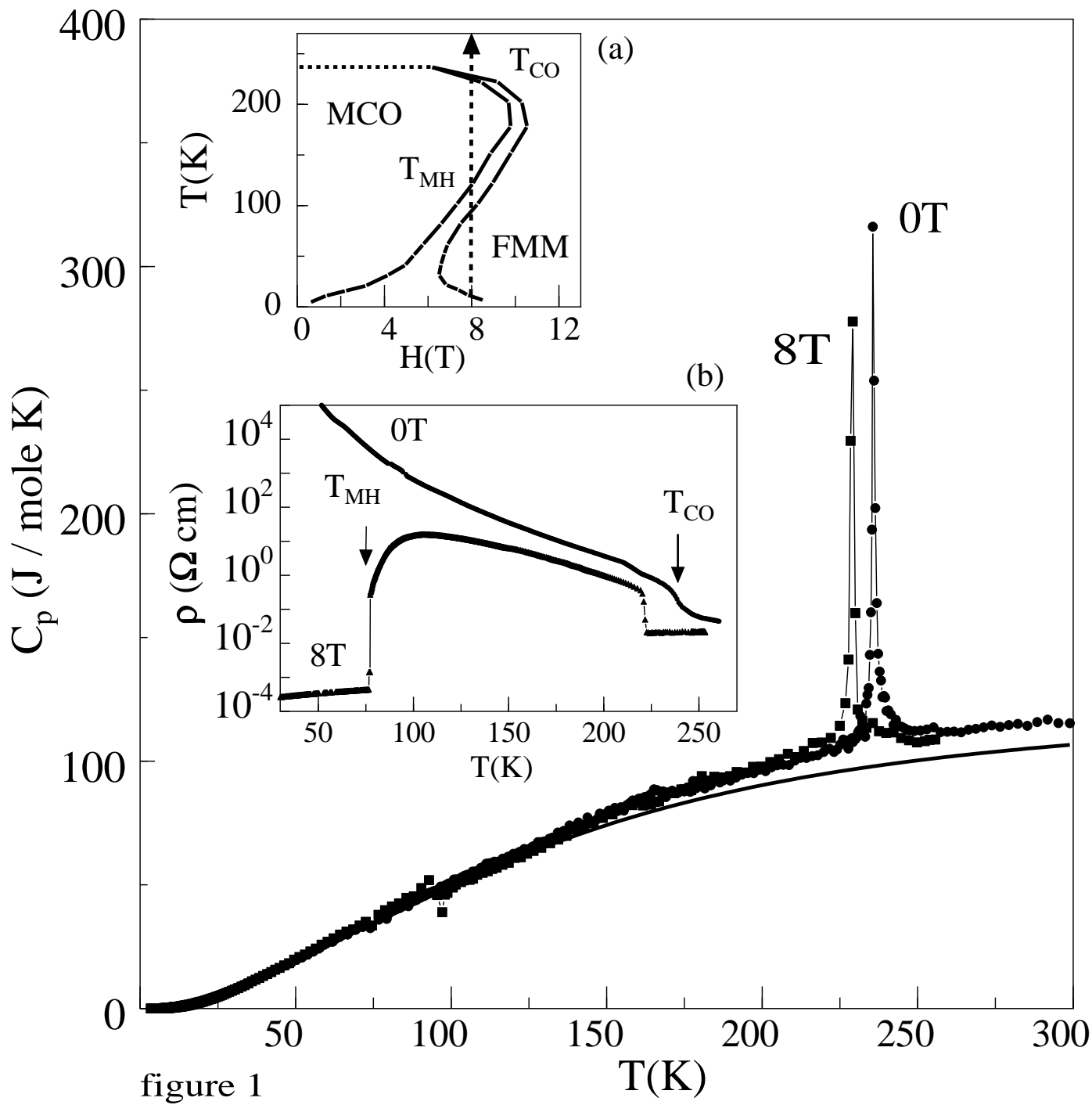
\end{document}